# Thermalisation and hard X-ray bremsstrahlung efficiency of self-interacting solar flare fast electrons

R. K. Galloway[1]*, P. Helander[2], A. L. MacKinnon[1], and J. C. Brown[1]

[1] Department of Physics and Astronomy, Kelvin Building, University of Glasgow, Glasgow G12 8QQ, U.K.
   e-mail: ross@astro.gla.ac.uk,a.mackinnon@educ.gla.ac.uk,john@astro.gla.ac.uk
[2] Max Planck Institute for Plasma Physics, Wendelsteinstrasse 1, 17491 Greifswald, Germany
   e-mail: per.helander@ipp.mpg.de

**ABSTRACT**

*Context.* Most theoretical descriptions of the production of solar flare bremsstrahlung radiation assume the collision of dilute accelerated particles with a cold, dense target plasma, neglecting interactions of the fast particles with each other. This is inadequate for situations where collisions with this background plasma are not completely dominant, as may be the case in, for example, low-density coronal sources.
*Aims.* We aim to formulate a model of a self-interacting, entirely fast electron population in the absence of a dense background plasma, to investigate its implications for observed bremsstrahlung spectra and the flare energy budget.
*Methods.* We derive approximate expressions for the time-dependent distribution function of the fast electrons using a Fokker-Planck approach. We use these expressions to generate synthetic bremsstrahlung X-ray spectra as would be seen from a corresponding coronal source.
*Results.* We find that our model qualitatively reproduces the observed behaviour of some flares. As the flare progresses, the model's initial power-law spectrum is joined by a lower energy, thermal component. The power-law component diminishes, and the growing thermal component proceeds to dominate the total emission over timescales consistent with flare observations. The power-law exhibits progressive spectral hardening, as is seen in some flare coronal sources. We also find that our model requires a factor of 7 - 10 fewer accelerated electrons than the cold, thick target model to generate an equivalent hard X-ray flux.
*Conclusions.* This model forms the basis of a treatment of self-interactions among flare fast electrons, a process which affords a more efficient means to produce bremsstrahlung photons and so may reduce the efficiency requirements placed on the particle acceleration mechanism. It also provides a useful description of the thermalisation of fast electrons in coronal sources.

## 1. Introduction

Most studies of the behaviour of fast electron populations in solar flares have assumed the presence of an ambient cold background plasma of sufficient density that it dominates the evolution of the fast electrons. In this case, self-interactions between the members of the fast electron population may be neglected, and only interactions between the fast electrons and members of the background plasma are considered. Analytic treatments of this kind include Brown's (1971) original formulation of the cold, thick target problem, and other analytic approaches such as those of Vilmer et al. (1986), which considers the spatial and temporal evolution of fast electrons in a region of inhomogeneous magnetic field and plasma density, and Conway et al. (1998), which gives analytic expressions for the moments of the electron distribution function.

Galloway et al. (2005) detailed a treatment which also assumed a dominant background plasma, but which relaxed the traditional cold target assumption. However, they also revealed a situation where observations implied that the fast electron population might not be insignificant in comparison to the ambient plasma, i.e. the local density of the fast electrons is not small compared to the density of the background plasma electrons. Krucker et al. (2009) have reported a coronal hard X-ray (HXR) source in which the fast particle density approaches the plasma density. In this case, it would be desirable to consider also the self-interactions between the members of the fast electron population. As well as being an interesting modelling problem, such an approach will have some bearing on a central problem in flare physics. The existing cold, thick target flare paradigm envisages the flare hard X-ray yield being generated principally by interactions between fast electrons and an essentially stationary (i.e. low thermal speed) background. Such encounters between particles of substantially different speed are highly inefficient in generating bremsstrahlung radiation: the fast electrons lose around $10^4$–$10^5$ times more energy through long-range Coulomb collisions than through bremsstrahlung emission of X-rays by short-range interactions (e.g. Brown 1971).

Comparison of the energy emitted in the form of X-rays with the total flare energy release reveals that, under the cold, thick target approximation, a substantial fraction of the total flare energy (possibly as much as 50%) is manifested in accelerated electrons (e.g. Brown 1971; Lin & Hudson 1976; Saint-Hilaire & Benz 2005). This has immediate, important implications for the particle acceleration mechanism: as many as $10^{36}$ electrons must be accelerated to energies greater than 20 keV each second

---

(Hoyng et al. 1976). If the acceleration mechanism is some form of magnetic reconnection in a single site, the acceleration region is limited to very restricted spatial scales, with current sheets only a few kilometres wide or thinner. The very great flux of accelerated electrons then constitutes a number and density problem: accelerating such a large number of electrons in sub-second timescales in such a small region requires a highly effective and efficient accelerating mechanism, which at this time remains theoretically elusive. One proposed solution is that the acceleration does not occur in a single or a few sites, but is distributed over a large number of small reconnection sites throughout the flaring volume (e.g. Turkmani et al. 2005). This somewhat relaxes the constraint of very small acceleration region volume, and so lowers the implied density of fast particles in the accelerating region. However, the overall number problem remains formidable – essentially, the acceleration mechanism is required to accelerate almost all the electrons present in the corona above an active region to energies above 20 keV every second for the duration of the impulsive phase of the flare.

An alternative solution would be to reduce the total quantity of energetic electrons required to produce the observed bremsstrahlung X-ray yield. This might be done by re-cycling electrons repeatedly through the accelerator, so that each electron could produce many high energy photons (e.g. Brown et al. 2009). Alternatively, fewer electrons would be required if the bremsstrahlung emission process was more efficient than that envisaged in the cold target scenario. Since the fast electron population is assumed to be 'dilute' in the cold target case, the fast electrons only undergo electron-electron collisions with target plasma electrons of much lower energy. Thus, the fast electrons rapidly lose energy to the target, preventing them from generating further high energy photons through bremsstrahlung emission during interactions with the ions. However, if the high energy particles collide with each other, the collisions merely exchange energy between the fast particles, giving less systematic collisional energy loss from the energetic electron population. The fast electrons can therefore continue to generate high energy photons. Consequently, a population of fast particles thermalising through self-interactions would constitute a more efficient bremsstrahlung source, allowing the observed flare hard X-ray output to be reconciled with a smaller accelerated electron population. This in turn relaxes the stringent requirements on the effectiveness of the acceleration mechanism.

In view of this, we now examine an approach to obtaining an approximate analytic solution for the time evolution of a population of self-interacting fast particles. To render the problem more tractable, and to afford a completely analytic solution, we consider a situation with no background plasma electrons and only Coulomb collision interactions exchanging energy between the fast electrons, and with electron-ion encounters giving rise to bremsstrahlung emission. In doing so, we consider a situation more closely analogous to a coronal or loop-top source than a footpoint source. Masuda et al. (1994) first identified a distinct flare hard X-ray source at high altitude in the more tenuous coronal plasma, in addition to the usual footpoint sources located in the denser chromospheric plasma. A study by Petrosian et al. (2002) of loop-top emission in limb flares suggested that coronal HXR sources may be a common feature in all flares, and such sources have recently been reviewed by Krucker et al. (2008). In neglecting the dense chromospheric background plasma, our model is not intended to describe the fast electron behaviour at the footpoint sources, nor the sort of high ambient density coronal source described by Veronig & Brown (2004). However, it can be considered to approximately describe the evolution of a fast electron population in a low-density coronal source, where interactions with the background plasma will be less significant since the background density is several orders of magnitude lower than in the chromosphere, but the intensity of the fast electron beam remains unchanged. Recently, Krucker et al. (2009) have identified in RHESSI data a coronal HXR source that appears to have this character.

In Section 2 we outline the derivation of the time-dependent electron distribution function we obtain for this problem. In Section 3 we give some numerical illustrations of the behaviour of this solution, and in Section 4 we compare it to some observations of corresponding solar events. In Section 5 we compare the efficiency of our model to that of the traditional cold, thick target model for the production of hard X-ray photons. Our conclusions are summarised in Section 6.

## 2. Development of the distribution function for thermalising electrons

We consider a plasma with an initial power-law distribution of electron energies and cold (effectively stationary) ions, and study the evolution of the electrons towards a Maxwellian due to Coulomb collisions. The plasma is spatially homogeneous and thermally insulated (i.e. experiences no particle or conductive energy losses), and we neglect energy loss through radiation (justified below). For simplicity we consider only the electrons since the ions will remain cold throughout, and electron-ion collisions affect mainly the pitch angle of the electrons without substantially altering their energy (e.g. Trubnikov 1965). We assume that the distribution is initially isotropic so that it remains isotropic for all time and that almost all electrons are non-relativistic.

We will not attempt to construct an exact, self-consistent solution. Instead, we seek an approximate distribution function which is analytically accessible but still behaves correctly in the high and low energy regimes. The evolution of the parameters of the distribution and the relative magnitudes of its components will be constrained by the relevant conservation laws. More precisely, we will aim to describe the distribution $f(v, t)$ produced at time $t$ as a population of electrons evolves under binary collisions, given an initial distribution

$$f(\mathbf{v}, t = 0) = \frac{3(p-1)}{4\pi} n_{\text{tot}} \frac{v_0^{3p-3}}{\left(v^3 + v_0^3\right)^p} \tag{1}$$

The distribution function $f(v, t)$ has normalisation

$$\int f(\mathbf{v}', t) d^3 \mathbf{v}' = n_{\text{tot}}, \tag{2}$$

for all times $t$. Here $n_{\text{tot}}$ is the total electron number density. Equation (1) ensures that this normalisation is satisfied at $t = 0$. Because of its isotropy, the value of $f$ depends only on speed $v$.

The initial condition (1) resembles the kappa distributions found commonly in space plasmas (e.g. Maksimovic et al. 1997) in that it includes a 'core' with characteristic speed $v_0$ playing the role of a thermal speed, combined with a power-law tail at high energies. The kappa distribution in turn has been found to be consistent with the electron distributions present in coronal HXR sources (Kašparová & Karlický 2009). Its precise form allows us to carry out the subsequent discussion mostly analytically. As already noted for the kappa distribution by Kašparová & Karlický (2009), it possesses a maximum when rewritten as a distribution per unit energy and thus needs no independent invocation of a "low-energy cutoff". The power-law index $p$ we require to be greater than $5/3$ for finite total energy. $v_0$ acts as a lower 'turn-over', ensuring the distribution remains well behaved at $v = 0$ (cf. Brown & Emslie 1988).

We employ the Fokker-Planck formalism for particle evolution under binary collisions (e.g. Rosenbluth et al. 1957; Montgomery & Tidman 1964; Helander & Sigmar 2002). The kinetic equation is

$$\frac{\partial f}{\partial t} = \frac{L}{v^2} \frac{\partial}{\partial v} \left[ v^2 \left( \phi' f - \psi'' \frac{\partial f}{\partial v} \right) \right], \quad (3)$$

where

$$L = \frac{e^4 \ln \Lambda}{m_e^2 \epsilon_0^2},$$

and $f = f(v, t)$ is the electron velocity distribution function. The right hand side of Eq. (3) is the Fokker-Planck representation of the effect of electron-electron collisions.

The coefficients for drift ($\phi'$) and diffusion ($\psi''$) may be expressed in terms of the Rosenbluth potentials (Rosenbluth et al. 1957):

$$\phi(\mathbf{v}) = -\frac{1}{4\pi} \int \frac{f(\mathbf{v}')}{|\mathbf{v} - \mathbf{v}'|} d^3v',$$

$$\psi(\mathbf{v}) = -\frac{1}{8\pi} \int |\mathbf{v} - \mathbf{v}'| f(\mathbf{v}') d^3v'.$$

For suprathermal particles (i.e. $v \gg v'$), we make the approximation that $|\mathbf{v} - \mathbf{v}'| \approx v$ and take it out of the integral. For any monotonically decreasing $f$ (such as the power-laws we consider), this approximation will always become adequate for sufficiently large $v$: the integrand of $\phi$ does not diverge as $v' \to v$ because the surrounding volume element simultaneously tends to zero more rapidly than the term in the denominator (see also e.g. Spitzer 1956; Trubnikov 1965).

Using Eq. (2), the Rosenbluth potentials then reduce to

$$\phi(v) \approx -\frac{n_{\text{tot}}}{4\pi v},$$

$$\psi(v) \approx -\frac{n_{\text{tot}} v}{8\pi}.$$

Since here the second derivative of $\psi(v)$ is zero, the suprathermal particles experience no velocity diffusion (second term in Eq. 3) and undergo only systematic velocity change (first term in Eq. 3). The kinetic equation now takes the form

$$\frac{\partial \tilde{f}_\infty}{\partial s} = \frac{1}{u^2} \frac{\partial \tilde{f}_\infty}{\partial u}, \quad (4)$$

where $\tau$ is a collision time, defined by

$$\tau = \frac{4\pi \epsilon_0^2 m_e^2 v_0^3}{n_{\text{tot}} e^4 \ln \Lambda}. \quad (5)$$

and we have introduced dimensionless time $s = t/\tau$ and speed $u = v/v_0$. We have also made $f$ non-dimensional by defining

$$\tilde{f}(u, s) = \frac{v_0^3}{n_{\text{tot}}} f(u, s)$$

and added the subscript $\infty$ to emphasise that this solution holds strictly in the high-velocity limit. The characteristic time $\tau$ could be evaluated for any speed; the appearance of $v_0$ in the initial condition (1) makes this the natural choice.

It is easy to show that any function solely of the combination

$$\left( u^3 + 3s \right)^{1/3}$$

is a solution of Eq. (4). Thus, with initial condition (1), the distribution at high energies at some time $s > 0$ will be

$$\tilde{f}_\infty(\mathbf{u}, s) = \frac{3(p-1)}{4\pi} \left( 1 + u^3 + 3s \right)^{-p}, \quad (6)$$

Employing this solution for $f(v)$ in the full form of the Rosenbluth potentials does not result in an identically zero $\psi''$ as we have assumed. However, for realistic $p$ values, $\psi''(v)$ is proportional to a large negative power of $v$, and so will be small in the high $v$ regime of interest to us here. We therefore consider this solution to be adequate for our approximate treatment.

The distribution $\tilde{f}_\infty$ given by Eq. (6) is also the one that would develop from our initial condition in the limit of zero ambient temperature if there was background plasma present in the system. For X-ray bremsstrahlung purposes, fast electron evolution is normally calculated in exactly this limit (e.g. Brown 1971; Melrose & Brown 1976; see also Takakura & Kai 1966). $\tilde{f}$ will be very close to $\tilde{f}_\infty$ for large $u$ but the differences will become more and more significant as $u$ becomes comparable to the mean speed of the whole distribution. Ideally we would determine the exact form of $\tilde{f}$ from Eq. (3) but we know nonetheless that collisions drive the whole distribution towards an isothermal, Maxwell-Boltzmann form and that $\tilde{f}$ will attain this form more and more closely as $s \to \infty$. $\tilde{f}_\infty$ conserves neither electron energy, since all electrons lose energy monotonically, nor number, since it implies an unphysical, non-zero flux of electrons out of the system at $u = 0$. In view of all this we should capture most of the essential physics of $\tilde{f}$ by adding a Maxwell-Boltzmann component to $\tilde{f}_\infty$, writing

$$\tilde{f}(u, s) = \tilde{f}_\infty(u, s) + \tilde{f}_{MB}(u, s)$$
$$= \frac{3(p-1)}{4\pi} \left( 1 + u^3 + 3s \right)^{-p}$$
$$+ \frac{\tilde{n}_M(s)}{\left( \pi \tilde{T}(s) \right)^{3/2}} e^{-u^2/\tilde{T}(s)} \quad (7)$$

Here $\tilde{n}_M(s)$ and $\tilde{T}(s)$ are the density and temperature of the Maxwellian component at time $s$, normalised to $n_{\text{tot}}$ and to $mv_0^2/(2k)$ respectively. Clearly $\tilde{n}_M(0) = 0$. $\tilde{n}_M(s)$ and $\tilde{T}(s)$ will be determined for $s > 0$ by appealing to conservation of electron number and energy, as follows.

We rewrite Eq. (2) in dimensionless units:

$$4\pi \int_0^\infty \tilde{f}(u, s) u^2 du = 1 \quad (8)$$

Eq. (7) in Eq. (8) immediately implies $\tilde{n}_M(s)$:

$$\tilde{n}_M(s) = 1 - (1 + 3s)^{1-p} \quad (9)$$

Equation (9) describes the evolution over time of the density of the thermal component of the overall distribution: as electrons are removed from the high-energy component they are added the Maxwellian component, increasing its density, such that the total density of the system is conserved.

For this thermally insulated system, in which radiation and electron-ion equilibration are neglected, the total energy density

$$\mathcal{E} = 4\pi \int_0^\infty \frac{m_e v^2}{2} f(v, t) v^2 dv$$

must also be constant. We introduce

$$\begin{aligned}\tilde{\mathcal{E}} &= \frac{2\mathcal{E}}{m_e v_0^2 n_{\text{tot}}} = 4\pi \int_0^\infty u^4 \tilde{f}(u,s) \mathrm{d}u \\ &= (p-1)\frac{\Gamma(\frac{5}{3})\Gamma(p-\frac{5}{3})}{\Gamma(p)}\end{aligned} \quad (10)$$

The last equality follows because $\mathcal{E}$ is constant so it may be evaluated at $s = 0$ using the initial condition (1).

Using (7) we can write the total energy in terms of $\tilde{T}(s)$:

$$\begin{aligned}\tilde{\mathcal{E}} &= 4\pi \int_0^\infty u^4 \tilde{f}(u,s) \mathrm{d}u \\ &= 4\pi \int_0^\infty u^4 \left(\tilde{f}_\infty(u,s) + \tilde{f}_{MB}(u,s)\right) \mathrm{d}u \\ &= (1 + 3s)^{5/3-p}\tilde{\mathcal{E}} + \frac{3}{2}\tilde{n}_M(s)\tilde{T}(s)\end{aligned} \quad (11)$$

Substituting (9) in (11) and rearranging we find

$$\tilde{T}(s) = \frac{2\tilde{\mathcal{E}}}{3}\left[\frac{1 - (1 + 3s)^{5/3-p}}{1 - (1 + 3s)^{1-p}}\right] \quad (12)$$

For early times, i.e. $s \ll 1$,

$$\tilde{T}(s) \approx \frac{2\tilde{\mathcal{E}}}{3}\frac{p - 5/3}{p - 1} \quad (13)$$

so for $p > 5/3$ the temperature is well-behaved at $t = 0$, justifying our earlier disregard of its initial behaviour. Note that the initial temperature of the Maxwellian component is determined completely by the parameters of the initial power-law, namely its energy and number densities and power-law index. We are not free to specify $\tilde{T}(0)$ independently. Electrons join the Maxwellian component only as they slow down to $u = 0$. Energy is being lost from all electrons included in $f_\infty$, however, so $\tilde{T}$ needs to take a finite value as soon as $s > 0$, for energy to be conserved.

For late times, i.e. $s \gg 1$,

$$\tilde{T}(s) \approx \frac{2\tilde{\mathcal{E}}}{3} \quad (14)$$

so all the energy resides in the Maxwellian part of the distribution, i.e. the plasma has completely thermalised. From (12), (13) and (14) we see that the temperature of the Maxwellian component will increase monotonically as the plasma evolves, but only by a factor of order unity for any plausible value of $p$.

Thus, the overall, evolving plasma distribution is described by Eq. (7), with $\tilde{n}_M(s)$ given by Eq. (9) and $\tilde{T}(s)$ given by Eq. (12). The initial distribution is a power-law with a lower turn-over. This thermalises, passing through intermediate stages with a modified power-law and growing thermal component. Eventually the power-law component diminishes completely, and the plasma is described by a purely thermal distribution. Although not a complete description, this approximate treatment will be correct at high velocities and late times and clearly includes, at least qualitatively, the essential physics of the situation.

## 3. Numerical illustrations; bremsstrahlung radiation

In this section we show numerical examples of the complete distribution function $\tilde{f}(u, s)$ and the resulting bremsstrahlung photon spectra, check numerically that entropy increases monotonically with time and address the neglect of radiation losses.

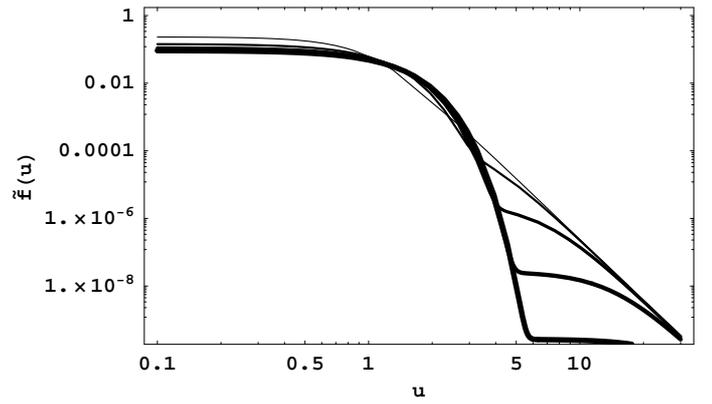

**Fig. 1.** Time evolution of the overall distribution function resulting from an initial power-law distribution with $p = 2$. Normalised evolution times of $s = 0.1, 10, 100, 1\,000$, and $10\,000$ are shown with progressively heavier lines.

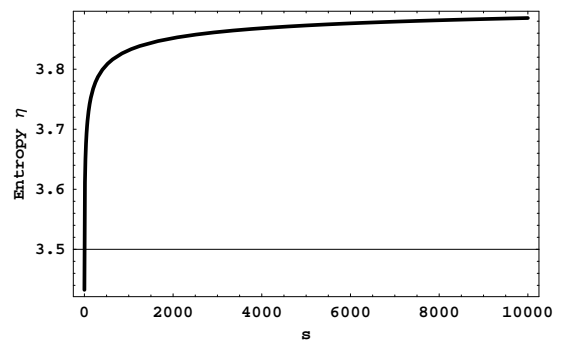

**Fig. 2.** Time evolution of the entropy of the overall distribution resulting from an initial power-law distribution with $p = 2$.

### 3.1. Distribution time evolution

The time evolution of the combined, overall distribution is shown in Fig. 1. As may be seen, the distribution begins as a power-law, flattening at low velocities. As time advances and the electrons in the power-law begin to thermalise, the distribution at lower energies takes on Maxwellian form, and the power-law tail diminishes. At late times, the Maxwellian is dominant and only a small power-law population remains, with a smooth intermediate transition to this regime from the initial condition.

### 3.2. Entropy

For Eq. (7) to be a valid solution, we have to check that the entropy increases with time. The entropy of the distribution is given by

$$\eta(s) = -\int \tilde{f} \ln \tilde{f} \mathrm{d}^3 u = -\int_0^\infty \tilde{f}(u,s) \ln \tilde{f}(u,s) 4\pi u^2 \mathrm{d}u. \quad (15)$$

Figure 2 shows a plot of the entropy of the electron population as a function of time for our example initial population with a power-law index of 2. As may be seen, the entropy does increase monotonically with time. Similar investigations of the time-dependence of the entropy for power-laws ranging from just greater than 5/3 (the boundary of validity of the solution) to 10 (steep), show that the entropy increase is smooth and monotonic over a range of appropriate power-law indices. Thus the solution does satisfy the increasing entropy criterion.

### 3.3. Photon spectrum

The initial impetus for this work came from considerations of X-ray production efficiency. Is the temporal evolution of hard X-ray emission in this scenario consistent with observations? We consider now the bremsstrahlung spectrum which would be emitted by our evolving electron population.

Let $\epsilon$ denote photon energy and write

$$\tilde{\epsilon} = \frac{2\epsilon}{m_e v_0^2}$$

Following the usual (thin target) formalism (e.g. Brown 1971) we find that the rate of photon emission is per second per unit photon energy per unit volume of the source region is

$$j(\epsilon) = 4\pi n_{\text{tot}} \int_{\sqrt{2\epsilon/m_e}}^{\infty} v f(v,t) \frac{\text{d}\sigma}{\text{d}\epsilon} v^2 \text{d}v, \quad (16)$$

where $\text{d}\sigma/\text{d}\epsilon$ is the cross-section differential in photon energy (e.g. Koch & Motz 1959). Writing photon energy $\epsilon$ in units of $m_e v_0^2/2$ and

$$\frac{\text{d}\sigma}{\text{d}\epsilon} = \frac{2r_e^2}{m_e v_0^2} \frac{\tilde{\sigma}_0(\tilde{\epsilon},u)}{\tilde{\epsilon}}$$

where $r_e$ is the classical electron radius, we find

$$\tilde{j}(\tilde{\epsilon}) = \frac{4\pi}{\tilde{\epsilon}} \tilde{J} \int_{\sqrt{\tilde{\epsilon}}}^{\infty} u^3 \tilde{f}(u,s) \tilde{\sigma}_0(\tilde{\epsilon},u) \text{d}u \quad (17)$$

where the dimensionless combination

$$\tilde{J} = r_e^2 v_0 \tau n_{\text{tot}}$$

and

$$\tilde{j} = \frac{\tau}{n_{\text{tot}}} j$$

Numerically, $\tilde{J} = 6.1 \times 10^{-8} E_0^2$ where $E_0$ is the energy in keV of an electron of speed $v_0$.

Figure 3 shows a plot of bremsstrahlung spectra from our electron population with power-law index $p = 2$ (corresponding to a power-law in electron energy with a spectral index of $\delta = 2.5$), plotted for photon energies $1 < \tilde{\epsilon} < 100$, and for various values of $s$. We used the non-relativistic Bethe-Heitler cross-section (Koch & Motz 1959, expression 3BN(a)), for which

$$\tilde{\sigma}_0(\tilde{\epsilon},u) = \frac{16}{3}\frac{1}{137}\left(\frac{c}{v_0}\right)^2 \frac{1}{u^2} \log\left[\frac{u+\sqrt{u^2-\tilde{\epsilon}}}{u-\sqrt{u^2-\tilde{\epsilon}}}\right]$$

$$= \left(\frac{9.95}{E_0}\right)\frac{1}{u^2}\log\left[\frac{u+\sqrt{u^2-\tilde{\epsilon}}}{u-\sqrt{u^2-\tilde{\epsilon}}}\right] \quad (18)$$

This cross-section yields at least the shape of the spectrum adequately in the 10s of keV regime.

The photon spectra reflect the distribution function behaviour shown in Fig. 1, displaying a smooth temporal and spectral transition from the initial condition, with a straight, power-law spectrum, to late times, where the Maxwellian dominates the spectrum and the power-law component is minimal. The power-law component hardens progressively with time.

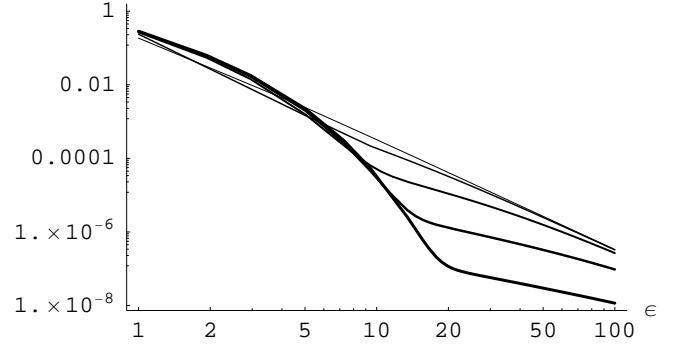

**Fig. 3.** Time evolution of the emitted bremsstrahlung photon spectrum resulting from the thermalisation of an initial power-law distribution with $p = 2$. Normalised evolution times of $s = 0.1, 10, 100, 1\,000$ and $10\,000$ are shown with progressively heavier lines.

### 3.4. Neglect of radiation losses

With Eq. (17) we may also check the neglect of radiation losses in this discussion. In dimensionless units, the total energy loss rate $R(s)$ to bremsstrahlung is

$$R(s) = \int_0^\infty \tilde{\epsilon} \tilde{j}(\tilde{\epsilon}) \text{d}\tilde{\epsilon}$$
$$= 4\pi \tilde{J} \int_0^\infty u^3 \tilde{f}(u,s) \int_0^{u^2} \sigma_0(\tilde{\epsilon},u) \text{d}\tilde{\epsilon} \text{d}u \quad (19)$$

where the last equality follows after substituting (17) and changing the order of integration. With cross-section (18) we may use the integral

$$\int_0^{u^2} \log\left[\frac{u+\sqrt{u^2-\tilde{\epsilon}}}{u-\sqrt{u^2-\tilde{\epsilon}}}\right] \text{d}\tilde{\epsilon} = 2u^2$$

to find that (19) becomes

$$R(s) = 8\pi \tilde{J}\left(\frac{9.95}{E_0}\right)\int_0^\infty u^3 \tilde{f}(u,s) \text{d}u \quad (20)$$

The integral in (20) gives the mean speed. It can be evaluated explicitly but it is enough to note that it varies only by a factor of order unity and is always $O(1)$ (just as $\tilde{T}$ is, as shown in Eqs. (12) - (14). The total dimensionless electron energy content (10) is also $O(1)$. If we take $E_0 = 10$ keV, $\tilde{J} = 5 \times 10^{-6}$ and we can see immediately, without any detailed time integration, that radiation losses will only become cumulatively significant on times $s$ of order $10^5$.

For low enough final temperatures atomic spectral lines will contribute substantially to radiation. If the final temperature is in the range $10^6 - 10^7$ K the total radiation loss could be up to an order of magnitude greater, depending on source chemical abundances (Sutherland & Dopita 1993). Even in this case radiation losses may be neglected at least up to $s = 10^4$ (independent of density, because $s$ is expressed in collision times).

Clearly a significant increase in HXR production efficiency is possible. In the next section we look more closely at whether the associated photon spectrum can reproduce observations.

## 4. Comparison with observations

In the previous section we see the characteristic spectral behaviour of the kind of source we model here: high-energy spec-tral hardening accompanied by the emergence of an isothermal spectrum at low photon energies. High-energy hardening ("soft-hard-harder" spectral behaviour) has been observed in a significant minority of flares (Frost & Dennis 1971; Cliver et al. 1986; Bai & Sturrock 1989; Hudson & Fárník 2002). Kiplinger (1995) found that around 15% of all flares display a soft-hard-harder spectral progression. Observations also exist of flares in which incoherent gyrosynchrotron emission (Melnikov & Magun 1998; MacKinnon 2006) is consistent with high energy electrons undergoing progressive hardening, but X-ray detectors lacked the required sensitivity at the relevant high energies to provide corresponding hard X-ray data for these events. This hardening is interpreted in terms of trapping of fast electrons in a low density, coronal region (e.g. Takakura & Kai 1966; Bai & Ramaty 1979). It occurs in our model for exactly the same reason and on its own would give no decisive test. The simultaneous emergence of the late phase, high temperature component (Fig. 3) would be much more suggestive, however.

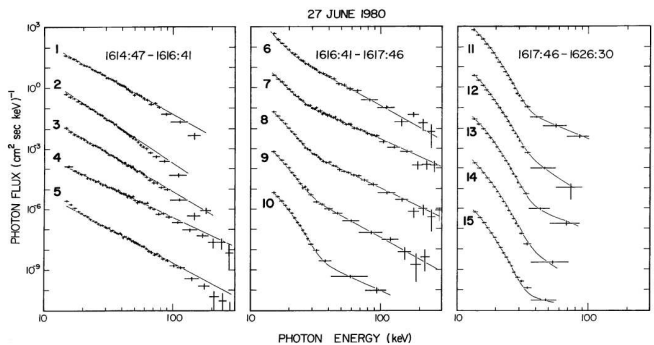

**Fig. 4.** Reproduction of Fig. 3 from Lin et al. (1981). Hard X-ray spectra from a range of time intervals during the flare of 27 June 1980. The absolute vertical scale corresponds to the uppermost spectrum in each panel. Each successive spectrum has been displaced downwards by two decades to enhance its visibility. However, the relative scalings of the spectra have been preserved. The solid lines are fits by Lin et al., and aid visual interpretation of the spectra. (Reproduced by permission of the AAS.)

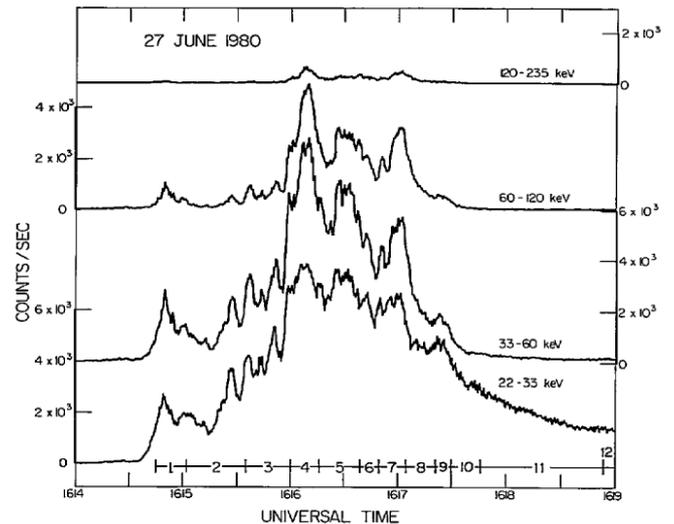

**Fig. 5.** Reproduction of Fig. 2 from Lin et al. (1981). Hard X-ray lightcurves for a range of energy channels during the flare of 27 June 1980. The marked time intervals correspond to those in the spectral plots in Fig. 4. (Reproduced by permission of the AAS.)

Lin et al. (1981) discovered a 'super-hot' (∼ 34 MK) thermal component of the hard X-ray spectrum emerging late in the impulsive phase of the 27 June 1980 flare. These observations were made with balloon-borne germanium detectors with 1 keV FWHM spectral response, which may be considered as precursors of the RHESSI detectors. The signature of such a super-hot thermal source had been seen before as a slowly-decaying emission component at ∼ 20–25 keV in scintillation counter detectors, but their spectral resolution was too low to confirm its identity as a thermal source. No imaging information was available for this event but such sources now appear to be a sub-class of coronal hard X-ray source (Krucker et al. 2008).

Figure 4 shows a summary of the observed X-ray spectra from this event for a set of 15 time intervals over the duration of the flare. These intervals are marked on Fig. 5, which shows the X-ray lightcurves for the event in a number of energy channels, as recorded by a scintillation counter also flown on the balloon. As may be seen, the spectra initially have a power-law form, appearing as a straight line in the log-log plots from 10 to 100 keV. As the event progresses, a departure from the straight power-law form may be seen, beginning in the lowest energies at interval 5. This departure grows increasingly prominent in later intervals, taking on a form which is well fitted by a Maxwellian component (Lin et al. 1981), and by interval 10 it is clearly the dominant component below 40 keV in the photon spectrum. The photon energy at which the spectrum departs from the straight power-law into the Maxwellian form increases with advancing time, rising from ∼ 20 keV at interval 5 to over 40 keV by interval 15.

Overall these observations show similar behaviour to the predicted bremsstrahlung spectra from our model, as shown in Fig. 3, with an initially power-law spectrum giving way to a growing Maxwellian component at low, but increasing, photon energies as the flare progresses. Given the approximate character of our treatment, we do not attempt detailed spectral fitting; but the qualitatively similar behaviour is clear.

In the model spectra, the mean intensity of the power-law, non-thermal component decreases with time. In addition, variation in its gradient (spectral hardness) may be seen. Taken over the whole spectrum, a broad characterisation of spectral hardness (as would be measured by low spectral resolution, pre-RHESSI instruments) would suggest that the spectrum softens, since the intensity at low energies is approximately constant but the intensity at high energies decreases. However, the portion of the spectrum which remains power-law in character (i.e. above $\epsilon \approx 20$) actually hardens, since lower energy electrons thermalise more readily. In a detector with sufficiently high spectral resolution, the power-law type portion of the spectrum could be separately identified and it would be seen to progressively harden. Such behaviour appears consistent with the recent RHESSI study of Shao & Huang (2009), for instance.

The intensity of the non-thermal component for the 27 June 1980 flare also decreases with time. However, the behaviour of its spectral hardness is less clear. As may be seen from Fig. 5, the overall flare event featured a number of hard X-ray bursts. (We may consider each burst as a separate instance of the electron thermalisation process we explore in this paper.) Lin & Schwartz (1987) conducted a detailed study of the power-law spectra from these bursts. They found that many displayed the more common soft-hard-soft spectral progression (e.g. Hudson & Fárník 2002; Grigis & Benz 2004), but some showed a progressive hardening.

A "super-hot" component is not seen so cleanly in many flare spectra. Krucker et al. (2008), however, identify such phenom-ena with coronal hard X-ray sources, consistent with the picture we develop here. Alexander & Metcalf (1997) conducted a detailed study of the Masuda et al. (1994) event as observed by the HXT instrument on the *Yohkoh* spacecraft (Kosugi et al. 1991). They characterise this coronal event as also showing the emergence of a high temperature (~ 40 MK) thermal component from an initially non-thermal source.

Figure 6 shows an example of four RHESSI spectra of a coronal source from various intervals over the duration of the occulted-footpoint flare of 4 April 2002 (Jiang et al. 2006). The two spectra in the upper panel of Fig. 6 show the initial power-law nature of the emission in the rise, or preheating, and impulsive phases of the flare. The lower panel shows the spectrum from approximately one minute after the impulsive phase observation, exhibiting a clear thermal component joining the hardening power-law. The lower panel also shows a spectrum from approximately two minutes later, in which the emission appears entirely thermal, with any remaining power-law component being lost in noise. The power-law spectral index undergoes progressive hardening over the duration of the first three intervals shown in Fig. 6, before showing an apparent abrupt softening between the third and fourth intervals (Jiang et al. 2006, Fig. 11). The temperature of the fitted thermal component rises between the second and third intervals, reaching a peak of approximately 25 MK (Jiang et al. 2006, Fig. 11), before beginning to cool over the remainder of the flare (by a cooling process which we do not attempt to model). Thus, these observations also show qualitative similarity to our model. Jiang et al. 2006 also discuss 5 other coronal sources in detail. All show the emergence of a thermal component from an initial power-law. The evolutions of the spectral indices of some of these flares are less clear: however, some are clearly soft-hard-soft in nature.

The observations of the 27 June 1980 flare shown in Fig. 4 span a period of approximately 12 minutes. Around 3 minutes elapse between interval 1, in which the spectrum appears purely power-law in nature, and interval 10, by which time the thermal component has become the dominant spectral feature. The coronal sources in the study of Jiang et al. (2006) feature an elapsed time of 2 to 3 minutes between the onset of the flare and observations showing the presence of a distinct thermal component. As may be seen from Fig. 3, the model bremsstrahlung spectra show a clearly defined thermal component after a typical elapsed (normalised) time of $s = 100$. On selecting a coronal density of $n_{tot} = 10^{10}$ cm$^{-3}$ (Jiang et al. 2006), and $v_0$ consistent with low-energy cutoffs of 6–12 keV (in line with the values employed in Galloway et al. 2005; see also Kane et al. 1992), a value of 100 for $s$ corresponds to real times of approximately 1 to 3 minutes, which are comparable to the evolution times of the observed flare electrons.

## 5. Efficiency of bremsstrahlung production

As noted above, the low radiative efficiency of the traditional, cold, thick target model leads directly to the high total energy content of flare nonthermal electrons. How much more efficient an X-ray source is the contained, self-interacting electron population? The previous section gives us some guidance as to the relevant parameter regime.

In the cold, thick target model the high-energy form $\tilde{f}_\infty$, given by Eq. (6) is assumed to hold exactly at all energies. The photon flux with this assumption is given by

$$\tilde{j}_{tt}(\tilde{\epsilon}) = \frac{4\pi}{\tilde{\epsilon}} \tilde{J} \int_{\sqrt{\tilde{\epsilon}}}^{\infty} u^3 \tilde{f}_\infty(u, s) \tilde{\sigma}_0(\tilde{\epsilon}, u) du \qquad (21)$$

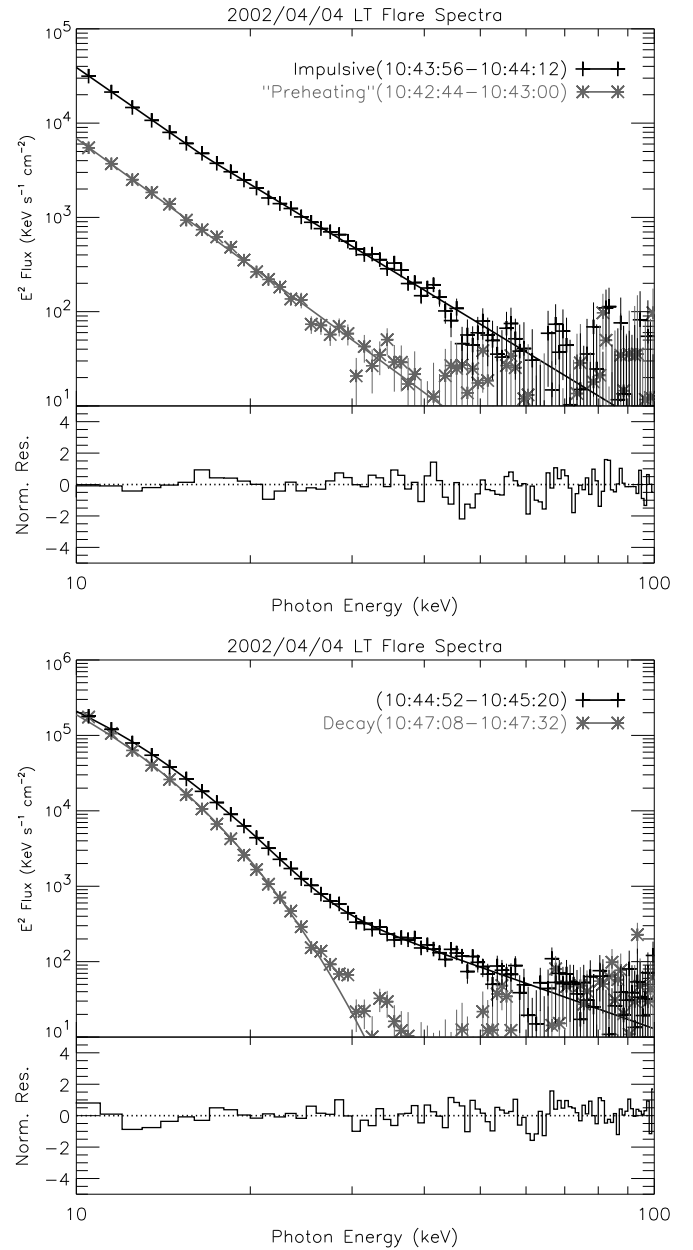

**Fig. 6.** Reproduction of Fig. 8 from Jiang et al. (2006). Hard X-ray spectra of the coronal source from four time intervals during the flare of 4 April 2002. The upper main panel shows the rise phase and the impulsive phase at the peak of HXR production. The lower main panel shows early and late decay phases. The smaller panels show the residuals (in units of the standard deviation) for thermal plus power-law model fits to the peak and early decay phase spectra. (Reproduced by permission of the AAS.)

In Figure 7 we compare the evolution in time of $\frac{d\tilde{j}}{d\tilde{\epsilon}}$ and $\frac{d\tilde{j}_{tt}}{d\tilde{\epsilon}}$ for photon energies $\tilde{\epsilon} = 4.5$, 6.75 and 9. Again we use the non-relativistic Bethe-Heitler cross-section. Multiplicative scaling constants depending on $v_0$ have been neglected since we wish only to compare the time behaviour in the two cases, at different photon energies. This temporal behaviour underlines the following findings.

Integrating $d\tilde{j}_{tt}/d\tilde{\epsilon}$ over time, from $s = 0$ to $\infty$, gives a finite result, the total thick target yield of photons at energy $\tilde{\epsilon}$ (Brown 1971). The relative efficiency of the self-interacting population

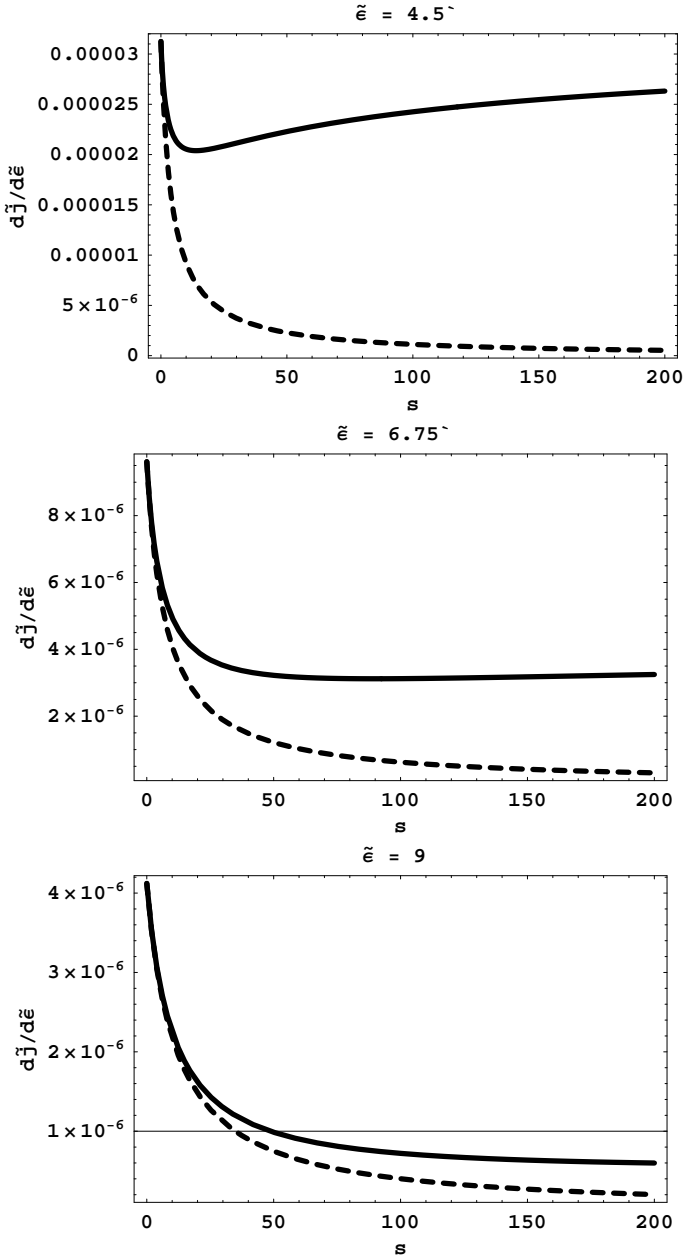

**Fig. 7.** Time evolution of the photon emission rates $\frac{d\tilde{j}}{d\tilde{\epsilon}}$ (self-interacting electrons: solid line) and $\frac{d\tilde{j}_{tt}}{d\tilde{\epsilon}}$ (cold, thick target: dashed line) for photon energies of (top to bottom) $\tilde{\epsilon}$ = 4.5, 6.75 and 9.

as an X-ray source may be described numerically by

$$\mathcal{R}(\tilde{\epsilon}_1, \tilde{\epsilon}_2, s) = \frac{\int_0^s \int_{\tilde{\epsilon}_1}^{\tilde{\epsilon}_2} \tilde{j}(\tilde{\epsilon}, s') d\tilde{\epsilon} ds'}{\int_0^s \int_{\tilde{\epsilon}_1}^{\tilde{\epsilon}_2} \tilde{j}_{tt}(\tilde{\epsilon}, s') d\tilde{\epsilon} ds'}$$

$$= 1 + \frac{\int_0^s \int_{\tilde{\epsilon}_1}^{\tilde{\epsilon}_2} \tilde{j}_{MB}(\tilde{\epsilon}, s') d\tilde{\epsilon} ds'}{\int_0^s \int_{\tilde{\epsilon}_1}^{\tilde{\epsilon}_2} \tilde{j}_{tt}(\tilde{\epsilon}, s') d\tilde{\epsilon} ds'}$$

(22)

where we have introduced $\tilde{j}_{MB}$ to stand for the photon flux from the Maxwell-Boltzmann component of $\tilde{f}$ on its own. From the second line of (22) we see immediately that the photon flux from the self-interacting electron population is always greater, in any photon energy range, than from the cold thick target.

$\tilde{n}_{MB}$ and $\tilde{T}$ and thus $\tilde{j}_{MB}$ tend to finite values as $s \to \infty$, so the total photon flux in any photon energy range may become arbitrarily large if we integrate over longer and longer times. The idealisation of a thermally isolated system thus complicates the comparison with the cold, thick target model.

With an appropriately chosen time interval, we can evaluate the relative X-ray efficiency of the self-interacting electron population by numerical evaluation of Eq. (22) (We used the Kramers cross-section $\tilde{\sigma}_0 \sim u^{-2}$ here because the triple integrals in the numerator and denominator of (22) may then be simplifed to single or double integrals, speeding up numerical evaluation. Since we only evaluate ratios of fluxes this should not result in any serious error). Taking the flare of 24 June 1980 (Lin et al. 1981) as a guide, we adopt $p = 2$ and fix $v_0$ by demanding that the final temperature is 24 MK, in agreement with observations. Then $m_e v_0^2/2 = 2.7$ keV. If we adopt for illustration $n_{tot} = 10^{10}$ cm$^{-3}$ then $\tau = 0.14$ s. If we then compare the fluxes over a period of 20 s, characteristic of impulsive phase hard X-rays (i.e. until $s = 140$), and over a photon energy range between $\epsilon_1 = 10$ keV and $\epsilon_2 = 100$ keV, then we find $\mathcal{R} = 7$. Extending the comparison to 30 s leads to $\mathcal{R} = 10$. The number of electrons needed would be smaller than that demanded by the cold thick target by a similar factor.

## 6. Conclusions

We have developed an approximate treatment of the thermalisation of a power-law population of electrons as a result of self-interactions in the absence of an ambient background plasma. The growth of the resulting thermal population is governed by the conditions of particle and energy conservation for the isolated plasma we consider. We have seen that our solution for the overall electron distribution function, as given by Eq. (7), satisfies the basic stipulations of the problem, i.e. a smooth transition from initial power-law to Maxwellian-dominated regimes, with relaxation occurring by collisions between the fast particles, and with monotonically increasing entropy. This solution corresponds to an evolving electron population which is more efficient in producing hard X-rays than the cold, thick target model. While a precise evaluation of the specific efficiency enhancement afforded by this process is hampered by some of the assumptions made in this treatment, we may nevertheless make an approximate comparison to the cold, thick target situation. We find that our model requires approximately a factor of 7 - 10 fewer accelerated electrons than the cold, thick target model to generate an equivalent hard X-ray photon flux. Thus, our model may alleviate some of the existing heavy requirements on the flare fast electron acceleration mechanism.

Eq. (10) predicts a simple relationship between the initial spectral index, measured at early times before the spectrum has thermalised, and the temperature of the thermal component that becomes dominant late in the evolution of the source. Both quantities measure the mean energy per particle and only conservation of total energy is needed for this relationship to be satisfied. With a large enough sample of sources this would provide a very good test of this picture.

More generally, many properties of the model are fixed by $p$ and $v_0$. The density is only needed to normalise to a given, total emission measure, and to determine the (dimensional) timescale for evolution.

The self-interacting source population's efficiency is limited in practice by the timescales for radiative or conductive energy loss, and by the validity of our assumption of perfect trapping. The latter may not be quite as unrealistic as it appears. Krucker et al. (2009) report the observation of an apparently contained source in which nonthermal electrons are dominant. A detailed comparison of our model with at least the decay phase of this event will be carried out in future work. Electrons accelerated at or near a magnetic null or region of very low field strength might be naturally contained near the acceleration site by the inevitably very high mirror ratios (Fletcher & Martens 1998).

A full evaluation of the validity of this approximate analytical treatment would depend on comparisons with a numerical solution using the complete expressions for the Rosenbluth potentials, which would be a non-linear problem. However, the suggestive similarity of the predictions of this initial model to some aspects of observed spectra indicates that a more complete treatment, in company with corresponding flare models, would be worth pursuing. Such an approach, incorporating additional properties such as energy losses from the thermal component, would facilitate a detailed, quantitative comparison between this model and the cold, thick target model. The consequences for the flare energy budget could then be investigated more fully. Such enhancements notwithstanding, these initial results nevertheless qualitatively reproduce the growth of a thermal component from a non-thermal, power-law HXR source over comparable timescales to those seen in some flare observations. Thus, they may provide useful insight into the evolution of flare electrons, particularly in coronal sources and those featuring a soft-hard-harder spectral progression.

*Acknowledgements.* During this work, R.K.G. was supported by a U.K. Particle Physics and Astronomy Research Council CASE Award and grant number PP/C000234/1. The work by P.H. was funded by the U.K. Engineering and Physical Sciences Research Council. Figures 4 & 5 and Fig. 6 by kind permission of R. P. Lin and Y. W. Jiang respectively and by the AAS. We thank Christina Burge for a careful reading of the manuscript and the referee for comments that resulted in improved presentation.